# Tailoring magnetic and hyperthermia properties of biphase iron oxide nanocubes through post-annealing

Supun B. Attanayake[1], Amit Chanda[1], Raja Das[2], Manh-Huong Phan[1], and Hariharan Srikanth[1,*]

[1]Department of Physics, University of South Florida, Tampa, FL 33620, USA

[2]SEAM Research Centre, South East Technological University, Waterford, Ireland

Tailoring the magnetic properties of iron oxide nanosystems is essential to expand their biomedical applications. In this study, the 34 nm iron oxide nanocubes with two phases consisting of $Fe_3O_4$ and $\alpha\text{-}Fe_2O_3$ were annealed for 2 hours in the presence of $O_2$, $N_2$, He, and Ar to tune the respective phase volume fractions and control the magnetic properties. X-ray diffraction and magnetic measurements were carried out post-treatment to evaluate the changes of the treated samples compared to the as-prepared, which showed an enhancement of the $\alpha\text{-}Fe_2O_3$ phase in the samples annealed with $O_2$, while the others indicated $Fe_3O_4$ enhancement. Furthermore, the latter samples indicated enhancements in the crystallinity and saturation magnetization while coercivity enhancement was most significant in the samples annealed with $O_2$, resulting in the highest specific absorption rates (up to 1000 W/g) in all the applied fields of 800, 600, and 400 Oe in agar during magnetic hyperthermia measurements. The general enhancement in the specific absorption rate post-annealing underscores the importance of the annealing atmosphere in the enhancement of the magnetic and structural properties of nanostructures.

*Corresponding author: sharihar@usf.edu

# 1. Introduction

The recent ability to fabricate magnetic structures of nanometer size, known as nanomagnetism, has allowed for the study of unique magnetic phenomena that are unobservable at larger length scales.[1–4] It has been shown that by utilizing nanostructures with varying sizes and shapes, their applicability in fields ranging from biomedicine to memory devices can be improved.[5–11] Iron oxide, one of the most historically studied magnetic systems has been found to be a unique platform to observe the interplay between different thermodynamically stable phases and their respective magnetic order, which has allowed for broad applications.[3,12–16] The already Food and Drug Administration (FDA)-approved composition is observed in many of the devices and has shown the potential to be tuned within its iron oxide composition to multiple magnetic phases with distinct features, thus the phase-tunability of iron oxide phases such as magnetite ($Fe_3O_4$), hematite ($\alpha\text{-}Fe_2O_3$), maghemite ($\gamma\text{-}Fe_2O_3$), Wüstite (FeO), etc. are currently being examined.[17,18] The phase-tunability of iron oxide nanostructures with size has shed light on the stability of certain phases in a variety of conditions, such as the FeO phase which possesses a higher stability in smaller structures of ~30 nm.[19] As an extension of an initial study of how the stability of iron oxide phases changes with size, different types of phase tunability in iron oxide nanostructures need to be explored. Prior research has shown that functionalizing iron oxide nanostructures for hyperthermia applications requires the tuning of their shape and size.[8,20] Of the various shapes that have been previously studied, nanocube structures of size ~30 nm have been shown to possess an elevated potential as compared to other structures such as spheres.[8] This is due to their enhanced anisotropy which results in higher specific absorption rates (SAR) and higher contrasts in magnetic resonance imaging (MRI) in comparison to its spherical counterparts.[7,8,20,21] Nature itself has shown the importance of contemplating iron oxide nanocubes with structures such as magnetosomes, found within magnetotactic bacteria which possess the capability to act as a compass utilizing the earth's magnetic field to navigate, opening up its use for further applications spanning from drug delivery to nanoelectronics.[22–24]

In this study, we explored the effects of annealing atmosphere to manipulate phase volume and hence magnetic properties of biphase iron oxide nanocubes for biomedical applications. Nanocubes consisting of $Fe_3O_4$ and $\alpha$-$Fe_2O_3$ with no observable exchange bias or superparamagnetic behavior were annealed in the presence of $O_2$, $N_2$, He, and Ar to observe compositional and magnetic changes. Herein, $O_2$ is a reactive gas that can directly interact with iron oxide structures while $N_2$, is a comparatively less reactive gas that can facilitate reaction-stimulating environments. The latter two gases are well-known inert gases with He being the lightest. Furthermore, the as-prepared and treated nanocubes were evaluated for their effectiveness in magnetic hyperthermia therapy, which showed that the coercivity enhancement plays a major role in tuning hyperthermia efficiency in comparison to the saturation magnetization enhancement.

## 2. Experiment

Synthesis of iron oxide nanocubes (NCs) was carried out following the procedures put forward by Nemati *et al*.[8] The chemicals Fe(III)-acetylacetonate (Fe(acac)$_3$, ≥99.9%), oleylamine (OA, 70%), oleic acid (OY, 90%), benzyl ether (BE, 98%), and 1,2-hexadecanediol (HDD, 90%) (Sigma-Aldrich, St. Louis, MO, USA) were uniformly mixed using a magnetic stirrer, in a three-necked round-bottomed flask with a jacketed heating mantle. The side openings were dedicated to a thermometer and continuous $N_2$ flow which was initially carried out with the condenser detached at the first stage, to facilitate degassing for 30 minutes at 110 $^0$C to ensure moisture and air are removed out of the flask. The condenser was then fitted with cold water circulation to ensure maximum retainment of the reactants throughout the process. Then the nucleation stage was initiated at 200 $^0$C, for 120 minutes with the final reflux stage carried out for 45 minutes, after which the solution is let to reach room temperature in the absence of a heat source. The obtained turbid reddish solution was then cleaned a minimum of two times with ethanol and a slight amount of hexane. The cleaning is carried out by back-to-back sonication and centrifuging of the solution, each time after the supernatant is disposed of. The obtained product was then let to dry to a fine powder which was evaluated for its structural and morphological characterization using a FEI Morgagni 268 transmission electron microscope (TEM) (FEI, Hillsboro, OR, USA) operating at 60 kV and was followed by compositional evaluation via diffractometry using a Bruker AXS D8 x-ray diffractometer (XRD) (Bruker, Madison, WI, USA) functional in Bragg-Brentano geometry at Cu K$\alpha$ wavelength. The obtained samples which conform to the initial structural and compositional evaluation were then separated, with some kept as controlled while the others were heated in the presence of multiple gases at 300 $^0$C for 2 hours in a ceramic combustion boat placed inside a tube furnace. The gases Oxygen ($O_2$), Nitrogen ($N_2$), Helium (He), and Argon (Ar) of ultra-high purity grade were obtained from nexAir. The treated samples were then again evaluated for their structural and morphological characterization using the TEM and XRD. All the magnetic measurements were performed in a DynaCool Physical Property Measurement System (PPMS) (Quantum Design, San Diego, CA, USA), utilizing the vibrating sample magnetometer (VSM) option. Magnetic hyperthermia measurements were performed by using a 4.2 kW Ambrell EasyHeat Li3542 system with varying magnetic fields (0-800 Oe) at a constant 310 kHz frequency. Samples were measured at 20 $^0$C as the starting temperature for a period of 300 seconds at 1 mg/mL nanoparticles in a 2% by-weight agar solution prepared using deionized water.

## 3. Results and Discussion

### *3.1. Structural Characterization*

The structural and morphological properties of the iron oxide nanocubes (NCs) were evaluated using a transmission electron microscope (TEM), which yielded a size distribution of nanocubes at 34 ± 4 nm. The as-prepared sample (AP) in the inset of Fig. 1(a) displays the uniform distribution of nanocubes which showed the presence of both ferrimagnetic (FiM) magnetite ($Fe_3O_4$) and antiferromagnetic (AFM) hematite ($\alpha$-$Fe_2O_3$) as seen in the X-ray diffractometry (XRD). When the AP samples were exposed to $O_2$, peaks related to $\alpha$-$Fe_2O_3$ further increased, indicating the $\alpha$-$Fe_2O_3$ volume fraction increment. The exposure to $N_2$ has removed a significant amount of the $\alpha$-$Fe_2O_3$ volume fraction, this change signifies the change of the oxidation number from 3 to 8/3, as the $\alpha$-$Fe_2O_3$ reduces to $Fe_3O_4$.[25] The reduction of the oxidation number confirms the ability of $N_2$ to act as a reducing agent at high temperatures and pressures.[25] The reaction, which has been reported for hematite nanowires at 350 $^0$C when annealed for 1 hour, seems to be initiated at 300 $^0$C in our iron oxide nanocubes when annealed for 2 hours.[26] Here both inert gases did not show significant phase changes from the AP. Though significant morphological changes were absent in all the samples post-annealing, a chain formation was observed, in the annealed samples is understood to be due to the diploe-diploe interactions between the NCs with the annihilation of the surfactants.[27] This chain formation can be advantageous for hyperthermia treatments as it tends to increase the specific absorption rate (SAR) values with the enhancement in anisotropic interactions.[23,24,28–30]

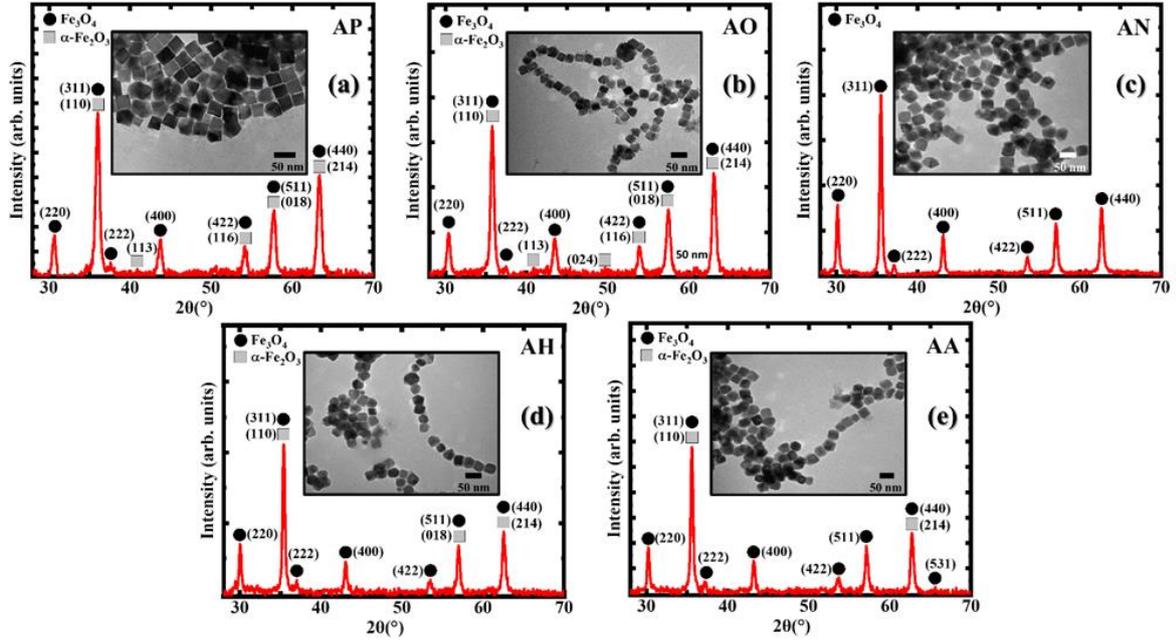

**Figure 1:** XRD patterns and the TEM images (inset) of (a) the as-prepared iron oxide nanocubes, (b) the iron oxide nanocubes annealed in O$_2$, (c) the iron oxide nanocubes annealed in N$_2$, (d) the iron oxide nanocubes annealed in He, and (e) the iron oxide nanocubes annealed in Ar.

*3.2. Magnetic Properties*

Temperature-dependent magnetization, *M(T)* measurements were carried out between 10-300 K following the zero-field cooled (ZFC), field-cooled cooling (FCC), and field-cooled warming (FCW) protocols in the presence of a 0.05 T magnetic field. The Verwey transition (VT), a first-order metal-insulator transition where the crystalline phase changes from high-temperature cubic to low-temperature monoclinic indicative of the Fe$_3$O$_4$ phase, is observed in all the samples except the sample annealed in O$_2$ (AO) between 109 K $\leq$ T$_V$ $\leq$ 121 K.[31–33] The sharpness of the transition which translates the crystallinity of the Fe$_3$O$_4$ phase can be observed to increase with annealing in the presence of N$_2$ (AN), He (AH), and Ar (AA), with the exception for the AO. The AO is understood to be oxidized in multiple stages, simultaneously with the final products of α-Fe$_2$O$_3$, following the reaction;[34,35]

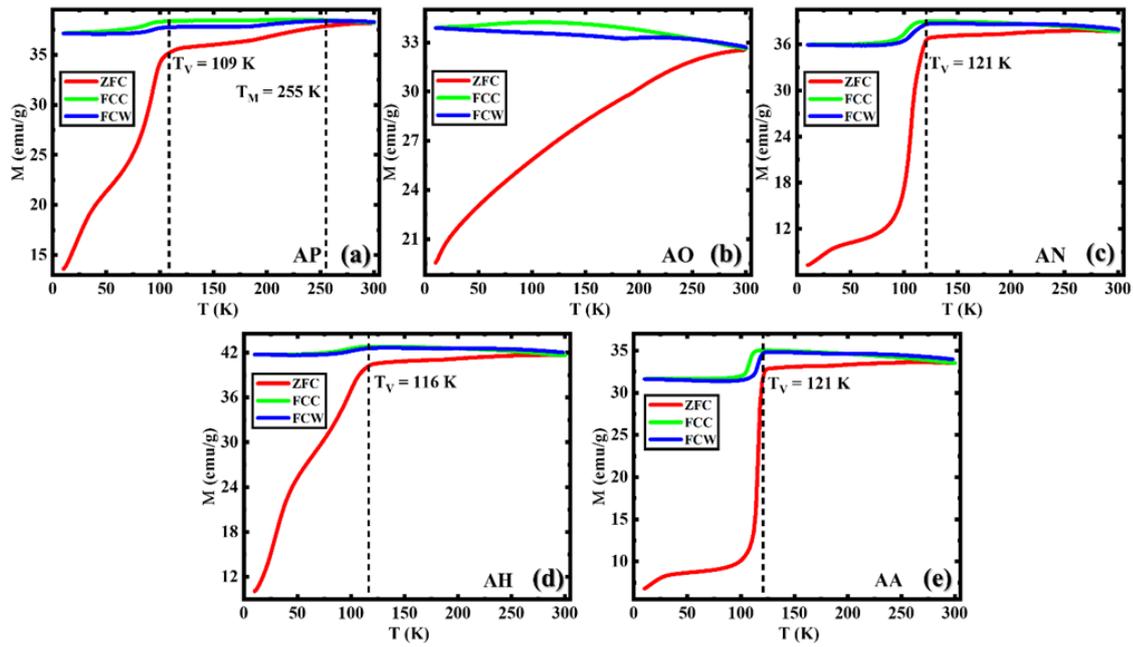

**Figure 2:** ZFC, FCC, and FCW M(T) curves measured in an applied field of 0.05 T for (a) as-prepared, (b) annealed in $O_2$, (c) annealed in $N_2$, (d) annealed in He, and (e) annealed in Ar.

The inside of the structure is expected to be oxidized to γ-$Fe_2O_3$-like (or γ-$Fe_2O_3$) phase from $Fe_3O_4$ which resulted in the disappearance of the VT.[34,36–39] The absence of the VT observed in prior annealing on iron oxide nanorods by Attanayake *et al*. supports the oxidation to γ-$Fe_2O_3$-like phase.[36,40] The AN, AH, and AA which indicated the reduction in the α-$Fe_2O_3$ phase indicated by the XRD data and the absence of the Morin transition (MT), are observed to show the transformation at comparative temperatures in the presence of various mixtures of gases, in our case $N_2$, He, and Ar.[41,42] Furthermore, AA indicates a slightly higher decrease in α-$Fe_2O_3$ which can be observed by the absence of (018) in Fig. 1(e) and the sharpness of the VT in Fig. 2(d) compared to AH.

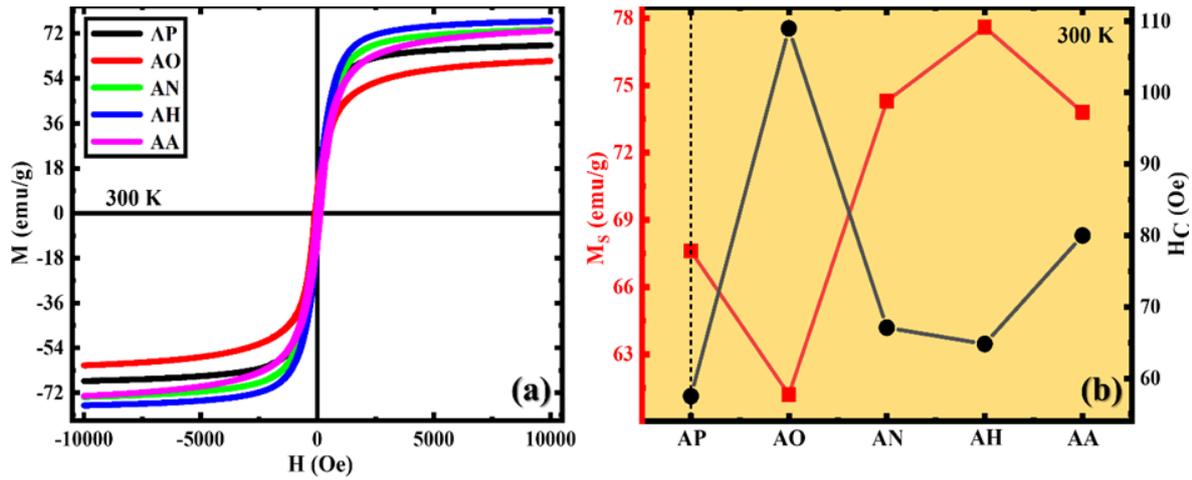

**Figure 3:** (a) Magnetic hysteresis loops M(H) of as-prepared, annealed in $O_2$, annealed in $N_2$, annealed in He, and annealed in Ar, and (b) variation of magnetization at the maximum value of the applied magnetic field ($M_S$) and coercivity ($H_C$) deduced from the M(H) loops.

The first-order magnetic transition, MT commonly known as the temperature-driven spin-flop transition is associated with the presence of the $\alpha$-$Fe_2O_3$ phase which occurs as the spin alignment changes from perpendicular to the c-axis above the $T_M$ to parallel to the c-axis below the $T_M$ is observable only in AP.[43–48]

The magnetization vs. applied field M(H) measurements shown in Fig. 3 were carried out at room temperature for a maximum 1 T applied field for all the samples. The magnetization at the maximum value of the applied magnetic field ($M_S$) decreased when moving from AP to AO but improved in all the other samples as the FiM $Fe_3O_4$ phase increased. Though the $M_S$ value decreased in AO, the value did not significantly decrease as observed with sole $\alpha$-$Fe_2O_3$ carrying nanostructures which lie less than 10 emu/g in most cases since the sample undergoes a two-step oxidation as explained by Zheng *et al*. which leaves a significant volume fraction of the FiM $Fe_3O_4$ phase intact.[34,49,50] The samples did not portray superparamagnetic features, similar to the observations made by Attanayake *et al.* but displayed an inversely proportional relationship between $M_S$ and coercivity ($H_C$), which can be related to the Stoner-Wohlfarth theory, $H_K = \frac{2K_{Ani}}{\mu_0 M_S}$ ; which shows that the magnetocrystalline anisotropy of a single domain particle is expected to show an inversely proportional relationship between $M_S$ and $H_K$, thus the $H_C$ change shown in Fig. 3b.[19,51,52] The general increment $H_C$ in all the annealed samples may be due to the magnetic hardening of the nanostructures with annealing, this is as the magnetic phases, especially the FM/FiM phases become more stable with annealing making them more resistant to demagnetization, and additionally defect formation with annealing can lead to magnetic pinning as it is a well-known defect engineering mechanism.[53,54] A $H_C$ of above 100 Oe was only registered in AO, understandably due to the combination of the effects; magnetic hardening, defect formation, and increased magnetic anisotropy.[7,54,55] Tunability of the AP in terms of $M_S$, and $H_C$ with varying gas types opens up the possibility for the nanostructures to be explored for the effectiveness of magnetic hyperthermia measurements with enhanced $H_C$ and $M_S$ separately, further tuned in accordance to varying applications such as magnetic particle imaging, magnetic tracers, etc..[53]

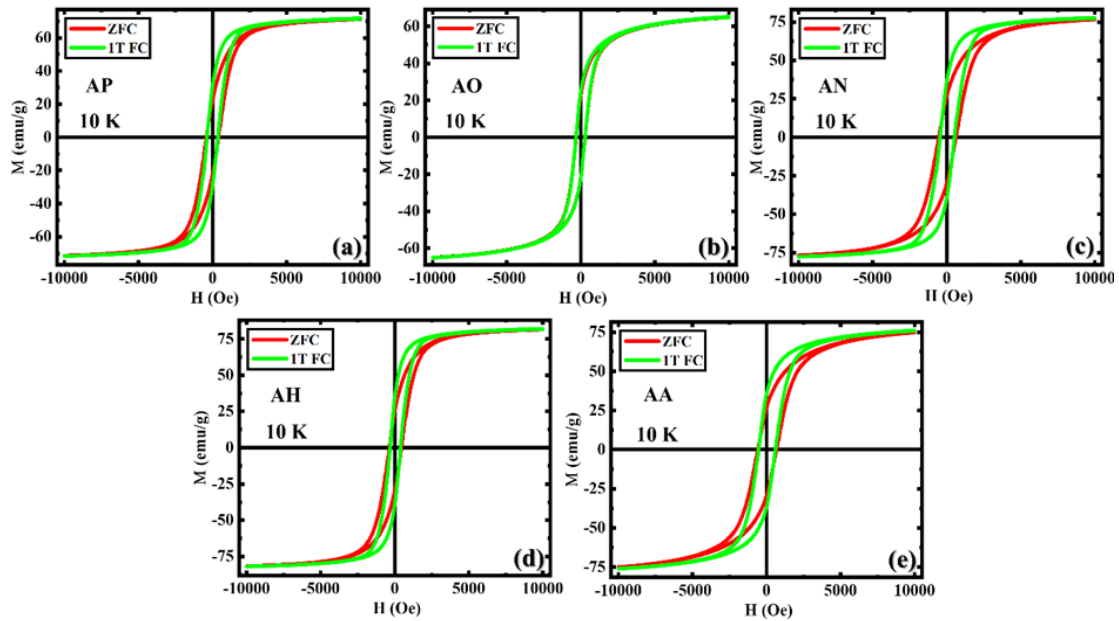

**Figure 4:** Magnetic Hysteresis loops M(H) recorded in ZFC and FC protocols in an applied field of 1 T for (a) as-prepared, (b) annealed in $O_2$, (c) annealed in $N_2$, (d) annealed in He, and (e) annealed in Ar.

Low-temperature *M(H)* measurements at 10 K showed a higher $M_S$ compared to the 300 K typical for nanostructures as the superficial spins tend to get better aligned with the applied magnetic field at lower temperatures as the thermal energy is lower, observed in smaller nanostructures with a high surface-to-volume ratio, which follows the Kneller's law which predicts an inversely proportional relationship between $H_C$ and T.[56,57] The ZFC and FC protocols applied on the nanostructures did not yield an exchange bias effect at the FiM $Fe_3O_4$ and AFM $\alpha$-$Fe_2O_3$ interfaces, understood to be due to the negligible interfacial interactions due to frozen spins.[19,37] The pinning effect is observed in all the samples and is the smallest in the AO, which may be due to the formation of a $\gamma$-$Fe_2O_3$-like phase similar to the observations made with iron oxide nanorods annealed with oxygen.[37]

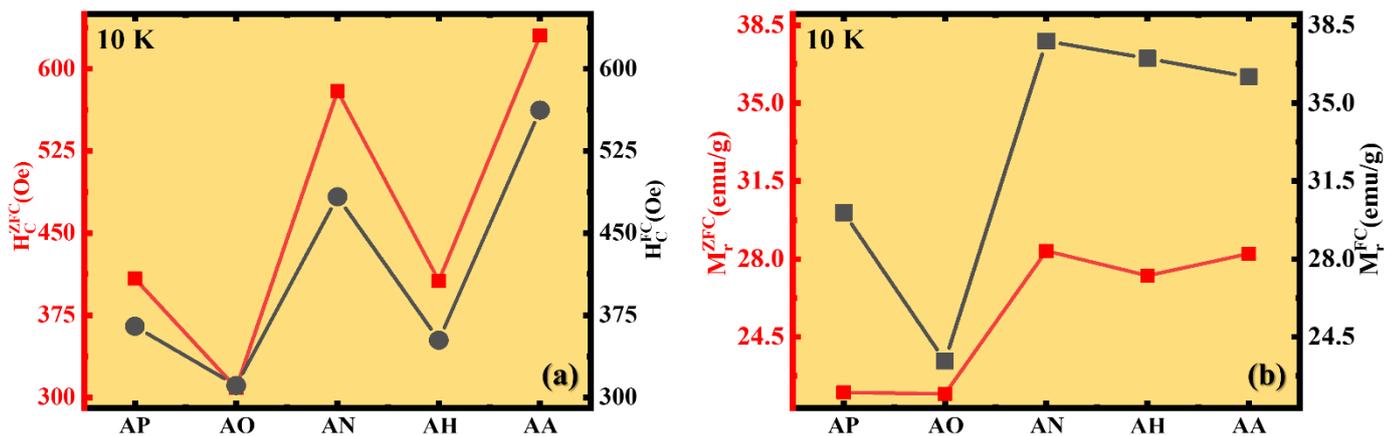

**Figure 5:** (a) $H_C$ values obtained at 10 K for ZFC and FC configurations, and (b) $M_r$ at 10 K for ZFC and FC configurations.

The $H_C$ with and without the application of the 2T field shows the separation with the observed biphases. The ZFC protocol which lets the spins be randomly oriented while being cooled down to the low temperature will possess a higher $H_C$ as the antiferromagnetic exchange coupling will pin the ferrimagnetic moments at the interface, but with the application of the FC protocol, the ferrimagnetic phase will align minimizing the interfacial pinning with the assistance of the cubic structure. The AO is the only sample that does not show a significant difference between the said two protocols with its dominant $\gamma$-$Fe_2O_3$-like phase, which does not possess the regular biphasic behavior in magnetic measurements. A similar observation is observed in the magnetization curves in Fig. 5(b) with AO.

3.1. Magnetic Hyperthermia

Figure 6(a-c) depicts the usage of samples in hyperthermia measurements, namely, the heating T(t) curves, at varying fields for a 300-second window, a relatively short period, which has been understood to be comparatively more effective in conjunction with radiation therapy at moderately lethal temperatures which are higher than the general hyperthermia window of 40-44 $^0$C.[58–61] Additionally, longer treatment periods are also associated with patient fatigue, adjacent healthy tissue destruction, and changes in the body conditions which can lead to complex reactions, adversely affecting the health and well-being of a patient.[62] Nanoparticle introduction was kept at 1 mg/mL, due to the tendency of instinctive reactions leading to adverse conditions, irrespective of the compatibility.[63,64] In the in-vitro study to ensure the performance of the nanoparticles in par or closer to the environment inside the body, a 2% by weight agar, a medium denser than water was used.[9,65] The SAR value of the heating curves was calculated using the initial slope method of the heating curve with the following equation;

$$SAR = \frac{\Delta T}{\Delta t}\frac{C_P}{\varphi}$$

The method devised, assumes a homogenous sample temperature and negligible heat loss at certain time intervals at the start of the alternating magnetic field.[66,67] Here the ΔT/Δt gives the rate of change of temperature, and $C_P$ indicates the specific heat capacity of the liquid medium. Here, the agar at 2% by weight is assumed not to have significantly changed its specific heat capacity and thus 4.186 J/(g $^0$C), the heat capacity of water is used. The symbol $\varphi$ indicates the mass of magnetic material per unit mass of liquid, which essentially gives the concentration used in the measurements.

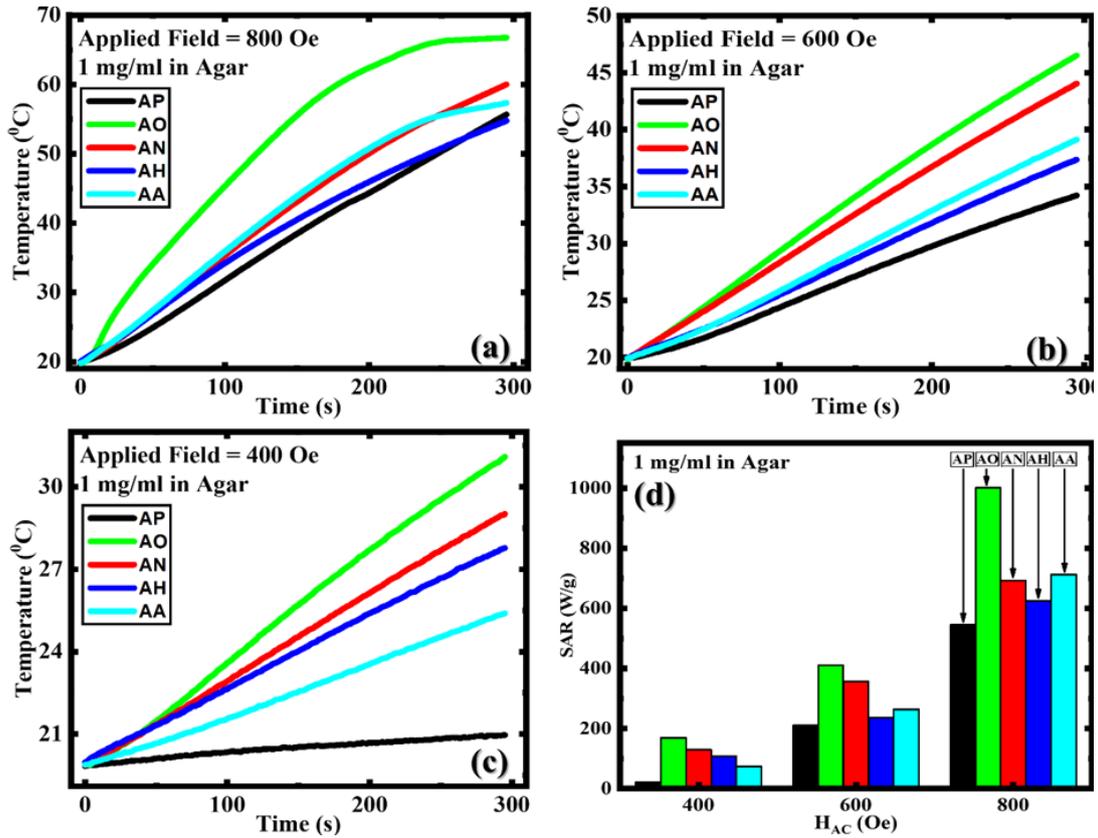

**Figure 6:** Heating curves of samples with concentration 1 mg/mL in agar, measured at 310 kHz frequency at (a) 800 Oe, (b) 600 Oe, (c) 400 Oe, and (d) SAR of all the samples measured at 400, 600, and 800 Oe.

It can be seen in Fig. 6d that all the samples in general showed significant improvement in heating potential and SAR values compared to the AP, while the AO showed an exceptional enhancement. The general enhancement of heating capabilities can be due to defect enhancement and increased grain size which in turn led to increased anisotropy.[53,68] The AO showed the highest SAR value in all the applied magnetic fields with the highest of 1001 W/g at 800 Oe (Fig. 6d, & **Table 1**), which is almost four times higher than that reported in FeO/Fe$_3$O$_4$ nanocubes synthesized by Khurshid *et al.*[21] Further when compared with the FiM iron oxide nanocubes of a similar size synthesized by Nemati *et al.* which showed almost the same magnetic properties compared to the AP sample (refer **Table 1**), the annealing was observed to have significantly increased the SAR in all the samples.[8] This suggests and re-affirms the already understood phenomena of SAR enhancement with the increment of $M_S$ and $H_C$.[69–71] In our study, the 34 nm nanocubes were used eliminating the variability of size and shape parameters. The adjusted annealing environments between the AO and AN samples clearly show that the $M_S$ and $H_C$ play a pivotal role in altering the SAR, but the enhanced $H_C$ in AO is observed to lead to a higher SAR value compared to the $M_S$ improved in the AN. Table 1 summarizes the change in $H_C$, $M_S$, and SAR in all the samples and highlights the exceptional SAR value observed in AO which possesses the highest $H_C$ underscoring the general relation between $H_C$ and SAR which supersedes the effect of $M_S$.

**Table 1:** Magnetic and Hyperthermia properties of the sample set at 300 K.

| Sample | $H_C$ (Oe) | $M_S$ (emu/g) | SAR in Agar @ 800 Oe (W/g) |
|---|---|---|---|
| *30 nm Nanocubes[8]* | *33.7* | *67.5* | *~540* |
| AP | 57.5 | 67.6 | 544 |
| **AO** | **109** | 61.2 | **1001** |
| AN | 67.1 | 74.3 | 692.8 |
| AH | 64.8 | **77.6** | 624.4 |
| AA | 80 | 73.8 | 712.3 |

## 4. Conclusion

The phase tunability of 34 nm iron oxide nanocubes with different types of gases enabled the enhancement of different fundamental magnetic properties, such as coercivity, saturation magnetization, and crystallinity. The enhancement of the specific absorption rate in all the samples with annealing, along with the general increment of coercivity reflected how magnetic anisotropy can affect magnetic hyperthermia efficiency. Comparing the annealing-induced changes in the coercivity and saturation magnetization, which are well-known parameters in tuning the specific absorption rate, it is observed that the former plays a larger role as the value in all the three tested applied magnetic fields were significantly higher in the AO, with enhanced $H_C$.

**Acknowledgments**

The research was supported by the US Department of Energy, Office of Basic Energy Sciences, Division of Material Science and Engineering under Award No. DE-FG02-07ER46438.